\documentclass[12pt,twocolumn]{aastex631}

\usepackage{natbib}
\usepackage{color}
\usepackage[english]{babel}
\usepackage[normalem]{ulem}
\usepackage{blindtext}
\usepackage{textgreek}
\usepackage{threeparttablex}
\usepackage{tabularx}
\usepackage{tabulary}
\usepackage{longtable}
\usepackage{rotating}
\usepackage{graphicx}
\usepackage{appendix}

\extrafloats{100}

\newcommand{\lsim}{\raise0.3ex\hbox{$<$}\kern-0.75em{\lower0.65ex\hbox{$\sim$}}}
\newcommand{\gsim}{\raise0.3ex\hbox{$>$}\kern-0.75em{\lower0.65ex\hbox{$\sim$}}}

\usepackage{listings}
\usepackage{xcolor}
\definecolor{codegreen}{rgb}{0,0.6,0}
\definecolor{codegray}{rgb}{0.5,0.5,0.5}
\definecolor{codepurple}{rgb}{0.58,0,0.82}
\definecolor{backcolour}{rgb}{0.95,0.95,0.92}
\lstdefinestyle{mystyle}{
    backgroundcolor=\color{backcolour},   
    commentstyle=\color{codegreen},
    keywordstyle=\color{magenta},
    numberstyle=\tiny\color{codegray},
    stringstyle=\color{codepurple},
    basicstyle=\ttfamily\footnotesize,
    breakatwhitespace=false,         
    breaklines=true,                 
    captionpos=b,                    
    keepspaces=true,                 
    numbers=left,                    
    numbersep=5pt,                  
    showspaces=false,                
    showstringspaces=false,
    showtabs=false,                  
    tabsize=2
}
\begin{document}

\title{The Delay Time Distribution of Quasi-Periodic Eruptions}

\author[0009-0005-1158-1896]{Margaret Shepherd}
\affil{Department of Astronomy, University of Illinois at Urbana-Champaign, 1002 W. Green Street, Urbana, IL 61801, USA}
\email{ms169@illinois.edu}

\author[0000-0002-4235-7337]{K. Decker French}
\affil{Department of Astronomy, University of Illinois at Urbana-Champaign, 1002 W. Green Street, Urbana, IL 61801, USA}

\author[0000-0001-9668-2920]{Jason T. Hinkle}
\affil{Department of Astronomy, University of Illinois at Urbana-Champaign, 1002 W. Green Street, Urbana, IL 61801, USA}
\affil{NSF-Simons AI Institute for the Sky (SkAI), 172 E. Chestnut St., Chicago, IL 60611, USA}
\affil{NHFP Einstein Fellow}

\author[0009-0008-7581-3096]{Ferdinand}
\affil{Department of Astronomy, University of Illinois at Urbana-Champaign, 1002 W. Green Street, Urbana, IL 61801, USA}

\author[0000-0001-9042-965X]{Samaresh Mondal}
\affil{Department of Astronomy, University of Illinois at Urbana-Champaign, 1002 W. Green Street, Urbana, IL 61801, USA}

\author[0009-0004-0436-0932]{Yashasvi Moon}
\affil{Department of Astronomy, University of Illinois at Urbana-Champaign, 1002 W. Green Street, Urbana, IL 61801, USA}

\author[0000-0003-1535-4277]{Margaret E. Verrico}
\affil{Department of Astronomy, University of Illinois at Urbana-Champaign, 1002 W. Green Street, Urbana, IL 61801, USA}

\begin{abstract}

Quasi-periodic eruptions (QPEs) are quasi-periodic X-ray bursts observed in the nucleus of a galaxy. Multiple pieces of observational evidence link QPEs to tidal disruption events (TDEs), which occur when stars are disrupted after approaching a supermassive black hole too closely. Post-starburst galaxies are overrepresented among the host galaxies of both TDEs and QPEs, though the mechanism causing this overrepresentation is unknown. While their physical origin is unclear, the delay time distribution (DTD) of QPEs, or rate of QPEs as a function of time since a burst of star formation, can constrain what mechanisms influence the QPE rate and possible QPE formation channels. We compile a catalog of 10 QPE host galaxies with optical spectra, model the stellar populations with \textsc{Bagpipes}, and retrieve the age of the most recent burst of star formation to construct the DTD of QPEs. We find that the QPE rate increases with post-burst age to reach a peak at $\sim$1 Gyr relative to a control sample, similar to the observational TDE DTD, though we cannot rule out a flat distribution of burst ages relative to a control sample. However, the fraction of QPE host galaxies with high ($>$1\%) burst mass fractions is larger than the fraction of galaxies with high burst mass fractions in either a sample of TDE host galaxies or a sample of control galaxies. If the preferred QPE formation channel requires extreme mass ratio inspirals (EMRIs), then such EMRIs may be more readily produced by large, $\sim$1-Gyr-old bursts of star formation.

\end{abstract}	

\section{Introduction}
\label{sec:intro}

Quasi-periodic eruptions (QPEs) are an emerging class of nuclear transients whose origin is not yet well understood but display a potential to expand our understanding of black hole environments and nuclear dynamics. QPEs are soft X-ray outbursts that occur on a roughly periodic basis, with periods ranging from hours to days, and emit from the nuclei of galaxies hosting a supermassive black hole \citep{Miniutti_2019, Giustini_2020, Arcodia_2021, Chakraborty_2021, Guolo_2024}. The duration and energy of QPE outbursts correlate with the period, displaying a duty cycle of $\approx$ 20\% \citep{Mummery_2025}. One model for QPEs suggests that the flares originate from interactions between two necessary ingredients: a disk and an extreme mass ratio inspiral (EMRI) object \citep{Dai_2010, Xian_2021, Franchini_2023, Linial_2023, HernandezGarcia_2025}. An EMRI is a compact object orbiting a supermassive black hole, gradually inspiraling via the emission of gravitational waves \citep{AmaroSeoane_2018}. \cite{Mummery_2025} investigates the nature of the EMRI, concluding the most likely scenario to be a star which has filled its Hills sphere, while \cite{Liu_2026} argue that the most likely EMRI to produce QPEs is a stellar-mass black hole. Alternative models also exist invoking disk instabilities \citep{Sniegowska_2020, Raj_2021}, EMRIs experiencing Roche lobe overflow \citep{Metzger_2022}, and partial tidal disruptions \citep{King_2020}.

Multiple pieces of evidence suggest that QPEs are linked to tidal disruption events (TDEs). A TDE is a type of transient event where a star that travels too close to a black hole is tidally disrupted by the black hole's tidal gravitational forces \citep{Rees_1988, Evans_1989, Gezari_2021}. TDEs have been found to be overrepresented among post-starburst (PSB) galaxies, which have recently had a burst of star formation but are no longer actively forming stars \citep{Arcavi_2014, French_2016, LawSmith_2017, Graur_2018, French_2020}. \cite{Wevers_2022} show that the incidence of QPEs in PSB galaxies and quiescent Balmer-strong (QBS) galaxies is much higher than the incidence of such galaxies in the local universe. \cite{Wevers_2024} show that extended emission line regions are overrepresented in both TDE and QPE host galaxies. \cite{Gilbert_2025} show that TDE and QPE host galaxies have similar S\'ersic indices, bulge-to-light ratios, and surface mass densities. Finally, several QPEs have occurred promptly following recent TDEs, on the order of $\sim$years after the TDE \citep{Nicholl_2024, Pasham_2024, Chakraborty_2025, Gilbert_2025, Baldini_2026}. 

A delay time distribution (DTD), or the rate of a transient as a function of time since a burst of star formation in the host galaxy, can be used to discern between different mechanisms that may impact the QPE rate. Comparing the QPE DTD and TDE DTD \citep{Shepherd_2026} will allow us to see if QPEs inherit the properties of TDE hosts or if other mechanisms dominate. In this work, we fit host galaxy spectra and recreate the star formation history (SFH) using the stellar population synthesis code Bayesian Analysis of Galaxies for Physical Inference and Parameter EStimation (\textsc{Bagpipes}, \citealt{Carnall_2018, Carnall_2019}). We construct an observational DTD of QPE host galaxies to extend the work of \cite{Shepherd_2026}. We describe the data in Section~\ref{sec:data}. Section~\ref{sec:methods} describes the methods, and Section~\ref{sec:results} describes the results. In Section~\ref{sec:discussion}, we discuss caveats and possible QPE formation channels. We conclude in Section~\ref{sec:conclusion}. Appendix~\ref{appendix:a} compares different SFH modeling attempts, and  Appendix~\ref{appendix:b} presents additional information on the QPE hosts galaxies.


\section{Data}
\label{sec:data}

\subsection{QPE Host Galaxy Spectra and Ancillary Data}
\label{sec:spectra}

We searched the literature for details on the QPEs that have been observed to date. We obtained optical host galaxy spectra via private communication, the Sloan Digital Sky Survey (SDSS, \citealt{Strauss_2002, Aihara_2011}), and archival data in the literature \citep{Hammerstein_2021, Evans_2023, Arcodia_2024}. In the case of QPEs that had occurred in galaxies that had recently hosted TDEs, we needed a pre-TDE spectrum or spectrum taken at least 365 days after the discovery of the TDE to avoid contamination. To balance the necessity of the largest sample size possible with sufficient data quality for stellar population fitting, we imposed a minimum signal-to-noise ratio (SNR) per pixel of 10\footnote{For galaxies with SDSS spectra, the SNR was pre-calculated. For galaxies with non-SDSS spectra, we calculated the SNRs in the wavelength band of $5200$ {\AA} $< \lambda < 5900$ {\AA} (rest wavelengths).}. When there were multiple host galaxy spectra for a single host, the spectrum with the highest SNR was chosen. 10 QPE host galaxies passed the selection cuts and were added to our sample\footnote{The host galaxies for AT2022upj and AT2019vcb were excluded due to available spectra being contaminated by their recent TDEs. The host galaxy for eRO-QPE5 was excluded due to available spectra not meeting the minimum SNR. The host galaxy for J2344 was excluded due to no available spectrum.}. We also collected the host galaxies' redshift and stellar mass (Table~\ref{tab:megatable}). The final QPEs in the sample are GSN069 \citep{Miniutti_2019}, RXJ1301 \citep{Giustini_2020}, eRO-QPE1 \citep{Arcodia_2021}, eRO-QPE2 \citep{Arcodia_2021}, eRO-QPE3 \citep{Arcodia_2024}, eRO-QPE4 \citep{Arcodia_2024}, XMMSL1J0249 \citep{Chakraborty_2021}, AT2019qiz \citep{Nicholl_2020}, Ansky \citep{HernandezGarcia_2025}, and SwJ0230 \citep{Guolo_2024}. 

\subsection{Control Samples}
\label{sec:control_samples}

In order to compare the QPE host galaxies to similar galaxies, we created a control sample matched in stellar mass and redshift to the QPE hosts. Using the same method as in \cite{Shepherd_2026}, we split the redshift range of the QPE hosts into two bins of size $\Delta z = 0.025$, and split the stellar mass range of the QPE hosts into three bins of size $\Delta \log_{10}(M_*/M_{\odot}) = 0.275$. We randomly select 100 non-QPE host galaxies with SNR $>$ 10 from SDSS that meet these requirements to be in the control sample. This provides a sample large enough for each QPE host galaxy to be compared to 10 field galaxies but small enough to run through \textsc{Bagpipes}.

When calculating the PSB overrepresentation, or the degree to which PSB galaxies are overrepresented in the QPE host galaxy sample compared to a comparison sample, we are not restricted by the need to run all of the comparison galaxies through \textsc{Bagpipes}; rather, we can collect the necessary information (H$\alpha$ equivalent width and Lick H$\delta_{\rm A}$ index) from SDSS. Thus, we created an expanded control sample using the same method as above. The size of this expanded control sample is 20,000 galaxies, or 2,000 galaxies per QPE host.

\section{Methods}
\label{sec:methods}

\subsection{\textsc{Bagpipes} Methods}
\label{sec:bagpipesmethods} 

\textsc{Bagpipes} \citep{Carnall_2018, Carnall_2019} uses the \textsc{MultiNest} nested sampling algorithm \citep{Feroz_2019} to estimate various galaxy parameters given a spectrum and priors provided by the user. To determine the DTD for QPE host galaxies, we need to know if and when a particular galaxy went through a burst of star formation. This requires us to provide \textsc{Bagpipes} with a functional form of the SFH. We enforce a two-component SFH as in \cite{Shepherd_2026}: an old stellar component modeled by a delayed exponential function ($SFR(t) \propto t \times e^{-t/\tau}$, \citealt{Simha_2014}) and a new stellar component modeled by a double power law function ($SFR(t) \propto [(t/\tau)^{\alpha} + (t/\tau)^{-\beta}]^{-1}$, \citealt{Wild_2020}) representing the burst. This approach allows for significant flexibility in the shape of the recent star formation history. This allows \textsc{Bagpipes} to fit galaxies with a recent burst, no burst at all, or ongoing star formation. See \cite{Shepherd_2026} for more details about the choice of SFH functional form (in particular their Appendix A), other priors provided to \textsc{Bagpipes}, and the full fit instructions. We calculated the time of observation to be the age of the universe at the redshift of the galaxy, using a flat cosmology defined by $H_0 = 70$ km s$^{-1}$ Mpc$^{-1}$ and $\Omega_M = 0.3$. Appendix~\ref{appendix:a} provides more detail about the dependence of \textsc{Bagpipes} results on the choice of SFH functional form.

\subsection{Measuring H$\alpha$ Emission and H$\delta_{\rm A}$ Absorption}
\label{sec:HalpHdelmethods}

To determine a host galaxy's type in the context of its SFH, we measure the equivalent width of the H$\alpha$ emission line and the Lick H$\delta_{\rm A}$ index in the galaxy spectra. Some of the QPE host galaxies had these values reported in the literature \citep{Wevers_2022}; for the other galaxies, we calculated these values ourselves\footnote{Though the spectrum used to analyze the host galaxy of eRO-QPE4 in \textsc{Bagpipes} due to its high SNR came from MUSE (Thomas Wevers, priv. comm.), we used a spectrum from SALT (Riccardo Arcodia, priv. comm.) to calculate its H$\delta_{\rm A}$ index due to its longer wavelength coverage.} (Table~\ref{tab:megatable}). We calculate the H$\alpha$ emission line equivalent width using \textsc{Bagpipes} (see \citealt{Shepherd_2026} for full details) and the Lick H$\delta_{\rm A}$ absorption line equivalent width using \textsc{pyLick} \citep{pyLick}. The galaxies in the expanded control sample had H$\alpha$ equivalent widths and H$\delta_{\rm A}$ indices available in SDSS {\tt galspec} \citep{Kauffmann_2003, Brinchmann_2004, Tremonti_2004}. We classify galaxies as quiescent, star forming, QBS, and PSB based on their spectrum's H$\alpha$ equivalent width and H$\delta_{\rm A}$ index using the same classification scheme as in \cite{Shepherd_2026}.

\section{Results}
\label{sec:results}

\subsection{QPE Rates versus Burst Age}
\label{sec:tderate_age}

\begin{figure}
\begin{center}
    \includegraphics[width=0.95\linewidth]{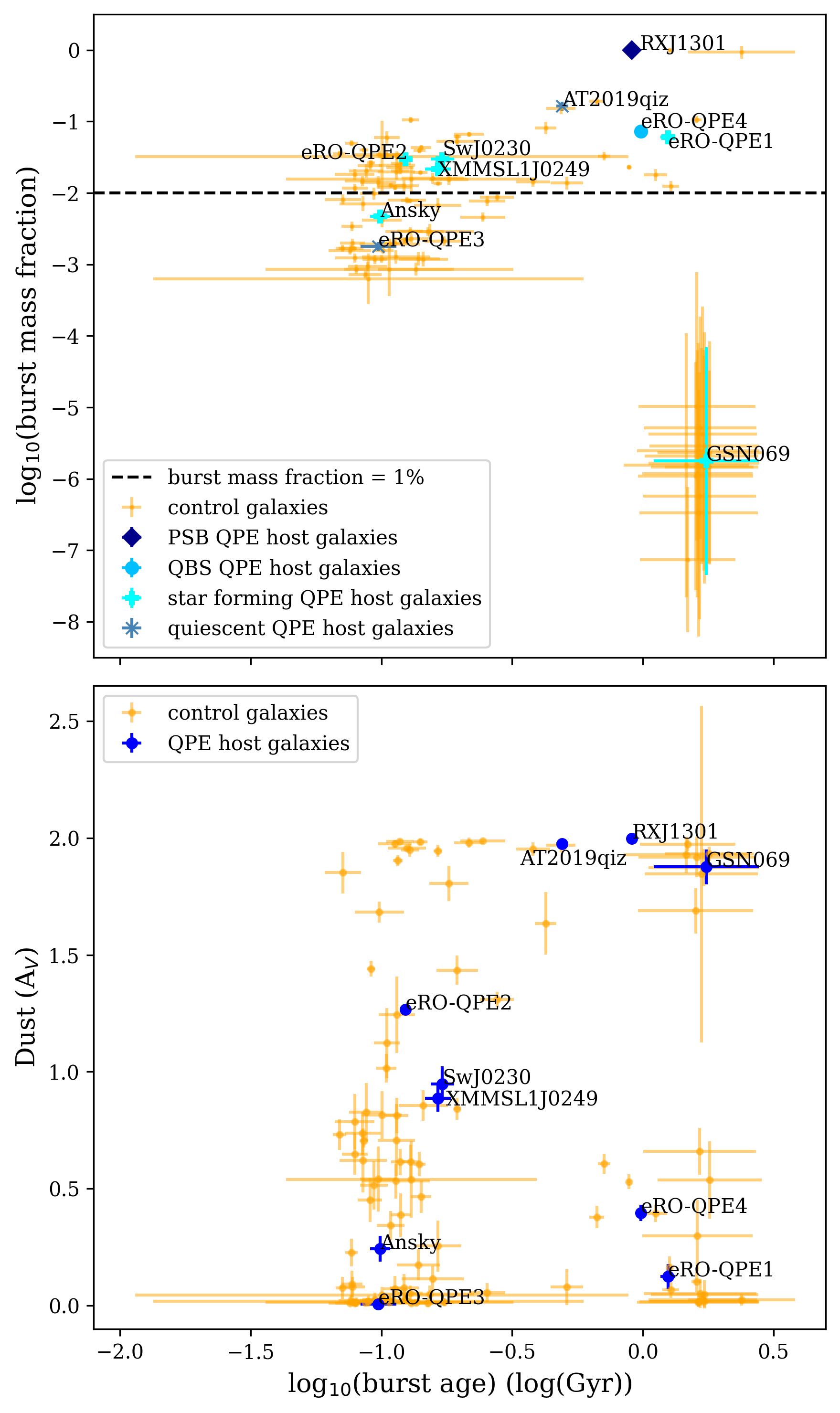}
\end{center}
\caption{{\it Top:} The burst mass fraction of the QPE hosts and the control galaxies versus the time since burst. The QPE hosts are shown in various shades of blue according to their galaxy classification (Section~\ref{sec:HalpHdelmethods}). The black dotted line indicates a burst mass fraction of 1\%, so the blue data points above that line are the QPE hosts that are considered to have a high burst mass fraction. {\it Bottom:} Dust attenuation of the QPE hosts and the control galaxies versus the time since burst. The data points corresponding to QPE hosts are labeled with the QPE names.}
\label{fig:bagpipes_results}
\end{figure}

We extract the posterior distributions of free parameters in the \textsc{Bagpipes} fit instructions. We report the results as the mean value and standard deviation of the relevant variables (Table~\ref{tab:megatable}). The burst ages for the QPE host galaxy sample and control sample are plotted against the burst mass fractions (stellar mass formed in the burst divided by total stellar mass formed) and the dust attenuation $A_V$ in Figure~\ref{fig:bagpipes_results}. Galaxies with low burst mass fractions tend to have large errors in their burst ages because it is uncertain whether or not a burst occurred at all. To select a subset of QPE hosts that have likely experienced a true burst of star formation, we create a subset of “high burst mass fraction” QPE hosts, whose burst mass fractions are over 1\%. This is the \textsc{Bagpipes} parameter we use to approximate a PSB-like SFH. The typical error in the burst ages of the high burst mass fraction QPE hosts is 5.5\%. The high burst mass fraction galaxies show a general trend of older bursts as the burst mass fraction increases. The potential impact of dust obscuration on reported burst ages is discussed in Section~\ref{sec:caveats_limits}.

The upper left panel of Figure~\ref{fig:finalrate_results} shows the cumulative distribution function of burst ages for the high burst mass fraction QPE hosts and the control galaxies. We use an Anderson-Darling two-sample test \citep{AndersonDarling} to determine if the sample of high burst mass fraction QPE hosts have a significantly different distribution of burst ages from the control galaxies. The resulting \textit{p}-value is 0.062 (with statistic = 1.74), indicating there is no significant difference between the burst age distributions of the two samples. 

The lower left panel of Figure~\ref{fig:finalrate_results} shows the rate and rate enhancement of QPEs as a function of burst age, also known as a DTD. To calculate the rate of QPEs as a function of time since burst, we use the burst ages of the high burst mass fraction QPE hosts, as well as the burst ages of the control galaxies to provide a normalization based on the distribution of burst ages in a non-QPE-hosting galaxy population. We calculate the QPE rate per post-burst age as
\begin{equation} 
R_{QPE,burst} =\frac{f_{QPE,burst}}{f_{con}} \times R_{avg},
\end{equation}
where $R_{QPE,burst}$ is the QPE rate per age bin in units of QPEs per year per galaxy, $f_{QPE,burst}$ is the fraction of high burst mass fraction QPE hosts that fall into each age bin, $f_{con}$ is the fraction of control galaxies that fall into each age bin, and $R_{avg} = 0.36 \times 10^{-4}$ QPEs per year per galaxy is the average QPE rate \citep{Arcodia_2024}. We calculate the QPE rate enhancement per post-burst age as
\begin{equation} 
\Gamma_{QPE,burst} =\frac{f_{QPE,burst}}{f_{con}}
\end{equation}
where $\Gamma_{QPE,burst}$ represents the QPE rate enhancement, or a multiplicative factor above the fiducial QPE rate.

The error bars on the DTD are Poisson errors using the \texttt{Pearson} method in the \texttt{astropy} function \texttt{poisson\_conf\_interval}, which are propagated through the rate and rate enhancement equations. We set the bin size by calculating $\frac{3\bar{\sigma_t}}{ln(10)\bar{t}}$, where $\bar{\sigma_t}$ is the average of the standard deviation values of the burst ages and $\bar{t}$ is the average of the burst ages. The dotted bars indicate the upper (1 $\sigma$) Poisson limit of the rate/rate enhancement for bins with no QPEs.

The overall trend of the DTD is that the QPE rate increases as a function of burst age to reach a peak at $\sim$1 Gyr, then decreases. However, due to the results of the Anderson-Darling two-sample test, we cannot distinguish this DTD from a flat DTD with no dependence of the QPE rate on burst age. Thus, this apparent peak at $\sim$1 Gyr is not significant, and we do not claim this DTD to be different from a flat distribution. The cumulative distribution function, in the top left panel of Figure~\ref{fig:finalrate_results}, can be used to see the unbinned distribution of burst ages of the control galaxies and the high burst mass fraction QPE hosts.

\subsection{QPE Rates versus Burst Strength}
\label{sec:tderate_strength}

\begin{figure*}
\begin{center}
\includegraphics[width=0.49\linewidth]{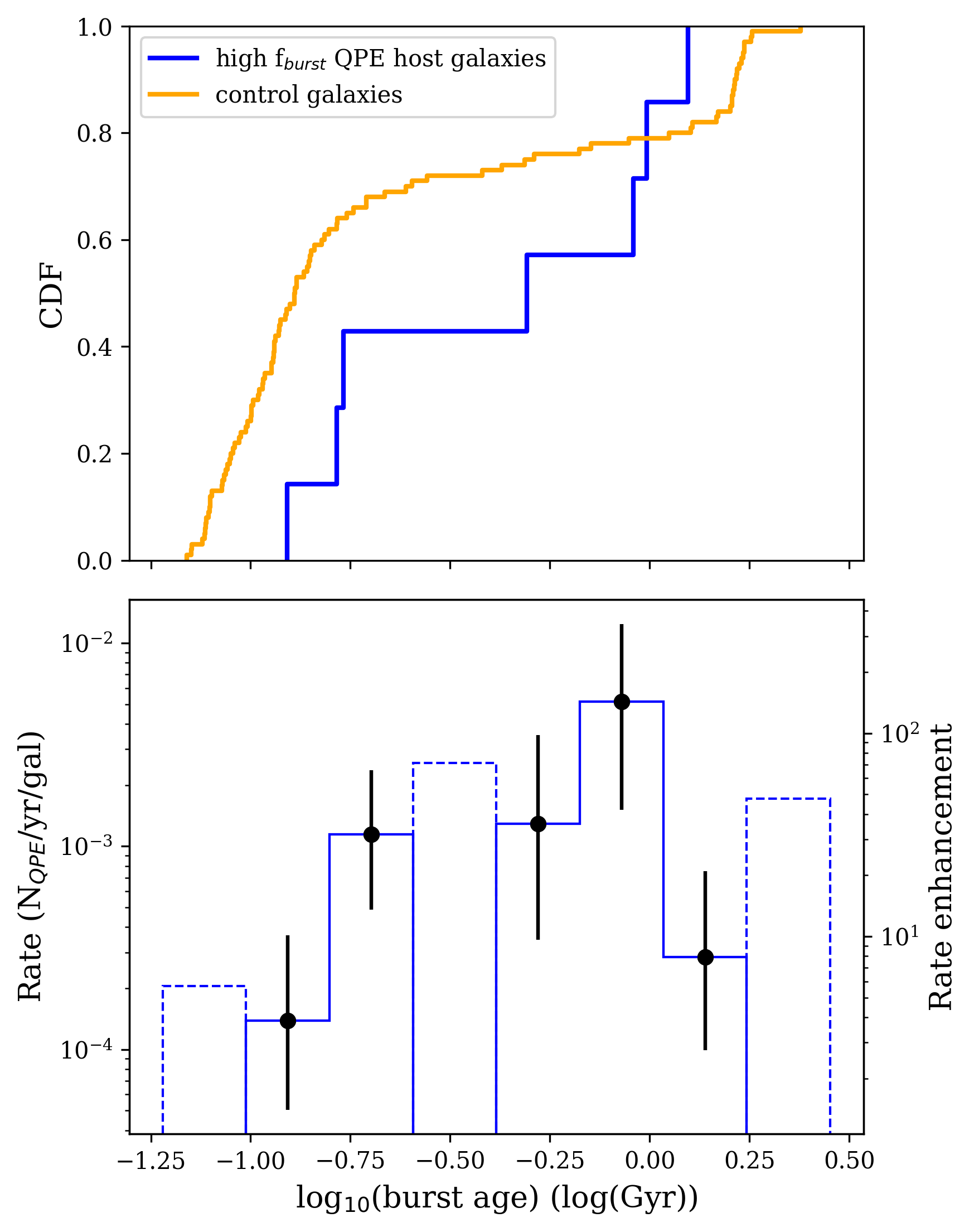}
\includegraphics[width=0.49\linewidth]{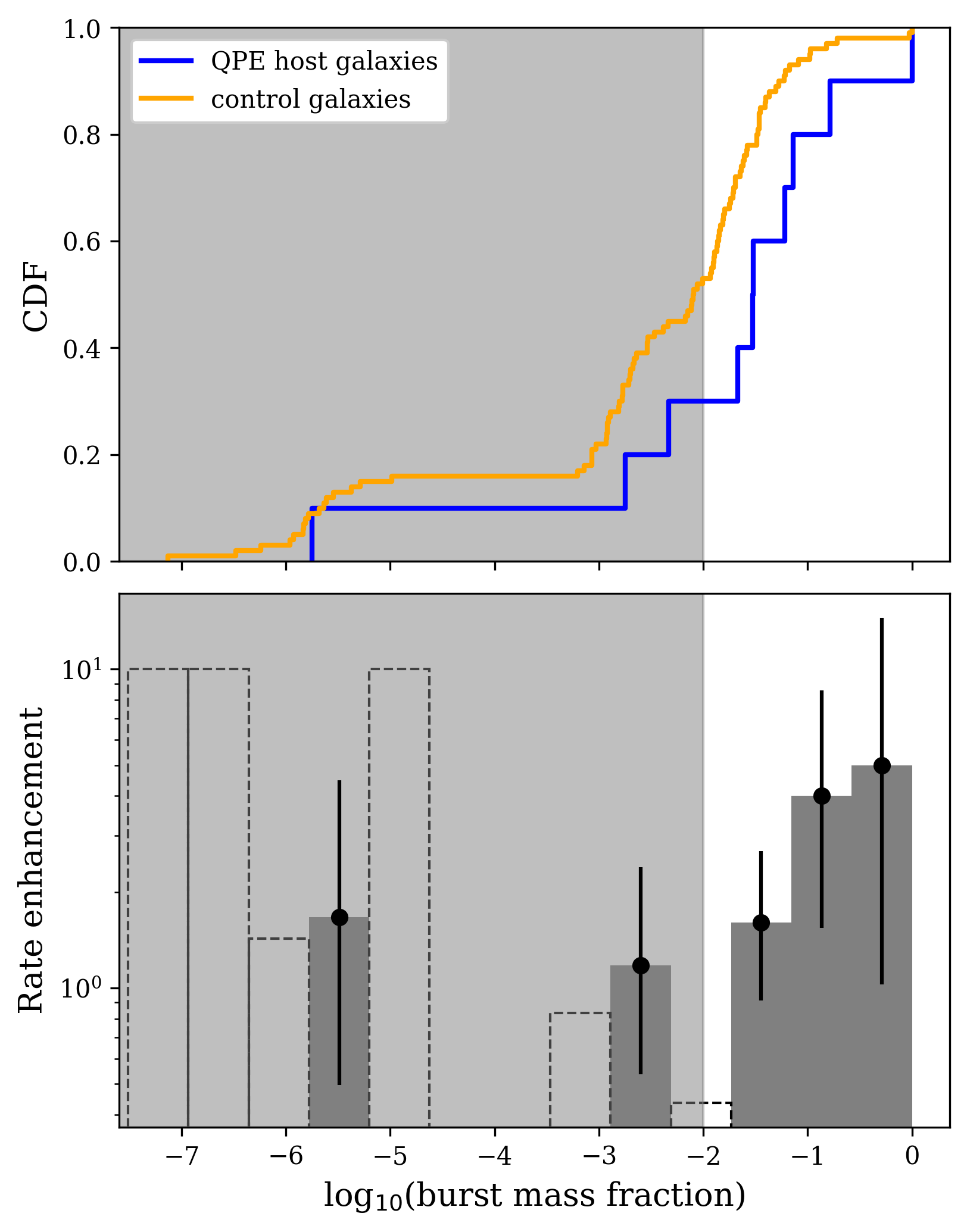}
\end{center}
\caption{{\it Top left:} A cumulative distribution function showing the burst ages of the QPE host galaxies that have had a significant burst of star formation and the entire control sample. {\it Bottom left:} The rate and rate enhancement of QPEs in galaxies that have experienced a significant burst of star formation as a function of burst age. {\it Top right:} A cumulative distribution function showing the burst mass fractions of the QPE hosts and the control sample. {\it Bottom right:} The rate enhancement of QPEs as a function of burst mass fraction of the host galaxy. The gray shaded regions indicates parameter space where large uncertainties in the burst mass fraction (see the y-axis on the upper panel of Figure~\ref{fig:bagpipes_results}) make interpretation difficult. In both bottom panels, the dotted histogram bars are bins that contain no QPE hosts but do contain control galaxies, and the height of the bin represents the upper limit on the rate/rate enhancement given the Poisson uncertainty on the number of QPE galaxies in that bin. There are no bars in bins where there are no QPEs and no control galaxies. The DTD shows a QPE rate/rate enhancement that increases with burst age, and the QPE rate enhancement versus burst mass fraction graph shows an increase in the rate enhancement at high burst mass fractions.}
\label{fig:finalrate_results}
\end{figure*}

We calculate the QPE rate as a function of host galaxy burst mass fraction as another way of investigating the relationship between the QPE rate and the host galaxy SFH. This is shown in the right panels of Figure~\ref{fig:finalrate_results}, where the upper panel shows the cumulative distribution function of the burst mass fractions of the QPE hosts and the control galaxies, and the bottom panel shows the rate enhancement of QPEs as a function of burst mass fraction. We use an Anderson-Darling two-sample test to determine if the QPE hosts have a significantly different distribution of burst mass fractions from the control galaxies. The resulting \textit{p}-value is 0.038 (with statistic = 2.26), indicating there is a significant difference. The burst mass fractions of the QPE host galaxies tend to be higher than those of the control galaxies. Additionally, the highest rate enhancement occurs at the highest burst mass fractions.

The shape of the bottom right panel of Figure~\ref{fig:finalrate_results} has some similarity to the analogous plot for TDE host galaxies \citep{Shepherd_2026}. The largest rate enhancement occurs when the burst mass fraction is at its greatest, for both QPEs and TDEs. However, there is also a substantial rate enhancement for QPEs at intermediate burst mass fraction (around $10^{-1}$), which is missing for TDEs. The TDE rate enhancement was concentrated at the highest burst mass fractions, while for QPEs the enhancement is present for a range of burst mass fractions greater than 1\% but is still highest at the highest burst mass fractions.

Seven out of the 10 QPE host galaxies in our sample, or 70\%, had high burst mass fractions, while 15 out of 42 TDE host galaxies in \cite{Shepherd_2026}, or 35.7\%, had high burst mass fractions. Using a binomial test, we determine whether or not it is significant that the proportion of QPE hosts with high burst mass fractions is higher than the proportion of TDE hosts with high burst mass fractions. We find (statistic = 0.7, \textit{p}-value = 0.029\footnote{We note that this may be considered to have marginal significance because we do multiple statistical tests in this paper, but the tests in this section show that we consistently find a difference (\textit{p}-value $<$ 0.05) when we compare QPEs' burst mass fractions to TDEs' burst mass fractions, and to the control sample burst mass fractions.}) that we would not expect 70\% or greater of QPE host galaxies to have high burst mass fractions if we assumed the rate of high burst mass fraction galaxies among TDE hosts to be standard.

\subsection{Post-Starburst Overrepresentation}
\label{sec:psb_overrepresentation}

\begin{figure}
\begin{center}
    \includegraphics[width=0.95\linewidth]{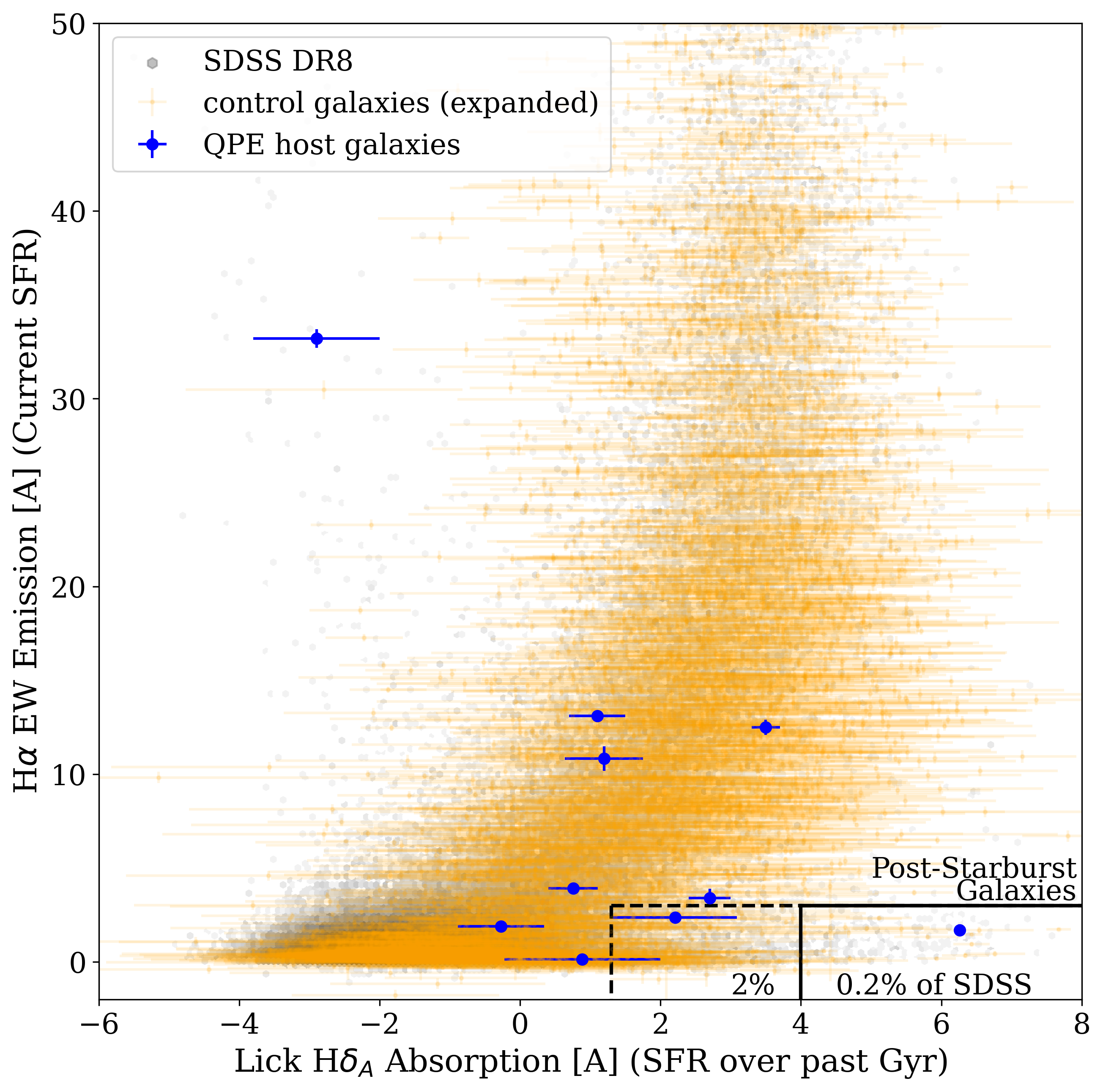}
\end{center}
\caption{The H$\alpha$ emission equivalent width and Lick H$\delta_{\rm A}$ absorption in the optical spectra of QPE host galaxies (blue points), galaxies from the expanded control sample (orange points), and SDSS DR8 (gray background). The box in the lower right corner bounded by solid black lines represents the region of parameter space where PSB galaxies reside, encompassing 0.2\% of galaxies in SDSS DR8. Extending the box towards lower H$\delta_{\rm A}$ values (dashed black lines) encompasses 2\% of galaxies in SDSS DR8, representing QBS galaxies. PSB and QBS galaxies are overrepresented in our QPE host galaxy sample compared to both SDSS DR8 and the expanded control sample.}
\label{fig:psb_overrepresentation}
\end{figure}

\cite{Wevers_2022} find that QPE hosts are overrepresented among QBS galaxies by a factor of 13, though their definition of QBS parameter space is slightly more generous than our definition\footnote{\cite{Wevers_2022} define QBS galaxies as having H$\alpha$ equivalent width $<$ 4 {\AA} and Lick H$\delta_{\rm A}$ index $>$ 1.31 {\AA}. This discrepancy results in this work having a different classification of eRO-QPE1 than \cite{Wevers_2022}.}. The H$\alpha$ equivalent width and the Lick H$\delta_{\rm A}$ indices of the QPE sample in this paper are shown in Figure~\ref{fig:psb_overrepresentation}. The x-axis displays absorption in the Lick H$\delta_{\rm A}$ index, a proxy for recent star formation due to the Lick H$\delta_{\rm A}$ index reaching maximum absorption when the dominant stellar population is A stars. The y-axis displays the equivalent width of the H$\alpha$ emission line, a proxy for ongoing star formation. PSB and QBS galaxies are located in the bottom right corner, indicating that they have very little ongoing star formation but have had significant star formation in the past Gyr. The gray points show galaxies in SDSS DR8 \citep{Aihara_2011}, while the orange points show the stellar mass-matched and redshift-matched control sample. The distribution of the QPE host galaxies in Figure~\ref{fig:psb_overrepresentation} differs from that of the comparison samples, especially in the low H$\delta_{\rm A}$ absorption-low H$\alpha$ emission and high H$\delta_{\rm A}$ absorption-high H$\alpha$ emission corners of the graph.

Using the percentage of galaxies in each comparison sample that are PSB and QBS, we can calculate the overrepresentation of PSB and QBS galaxies in our QPE host galaxy sample. Galaxies within H$\alpha < 3$ {\AA} and H$\delta_{\rm A} \geq 4$ {\AA} are PSB galaxies. 0.204$\pm$0.006\% of galaxies in SDSS DR8 fall in this regime \citep{French_2016}. However, 1 out of 10, or 10$\pm$9\%, of QPE host galaxies in our sample are PSB galaxies. The errors quoted are binomial errors. This results in an overrepresentation factor of 10/0.204 = 49.0$\pm$46.5 with respect to the background SDSS population. The expanded control sample discussed in Section~\ref{sec:control_samples} can be used to measure the overrepresentation factor while accounting for broader selection biases. PSB galaxies make up 73/20,000 = 0.37$\pm$0.04\% of the expanded control sample. This results in an overrepresentation factor of 10/0.37 = 27.4$\pm$26.2 with respect to the expanded control sample. 

Galaxies within H$\alpha < 3$ {\AA} and H$\delta_{\rm A}$ $\geq$ $1.3$ {\AA} are QBS galaxies. 2.32$\pm$0.02\% of SDSS galaxies fall in this regime \citep{French_2016}. However, 2 out of 10, or 20$\pm$13\%, of QPE host galaxies in our sample are QBS galaxies, including the PSB QPE host mentioned above. The overrepresentation of QBS galaxies among the QPE host galaxy sample is 20/2.32 = 8.6$\pm$5.4 relative to the background SDSS population. QBS galaxies make up 593/20,000 = 2.97$\pm$0.12\% of the expanded control sample. We calculate an overrepresentation value of 20/2.97 = 6.7$\pm$4.3 for the QBS galaxies with respect to the expanded control sample.

The overrepresentation numbers calculated above are plagued by large uncertainties due to the small number statistics of the QPE host galaxy sample, but are consistent with the trend observed among TDE host galaxies. In \cite{Shepherd_2026}, PSB (QBS) galaxies were found to be overrepresented among TDE host galaxies by a factor of 83$\pm$29 (14$\pm$3) when compared to galaxies in SDSS DR8. PSB (QBS) galaxies were found to be overrepresented among TDE hosts by a factor of 12$\pm$5 (8.6$\pm$2.2) when compared to a stellar mass-matched and redshift-matched control sample.

\section{Discussion}
\label{sec:discussion}

\subsection{Caveats and Limitations}
\label{sec:caveats_limits}

The overrepresentation of TDEs in PSB galaxies is present in optically/UV-selected TDEs \citep{Arcavi_2014, French_2016, LawSmith_2017, Graur_2018, Hammerstein_2021, Shepherd_2026}, but has not been shown in a sample of IR-selected TDE candidates \citep{Masterson_2024}. If the overrepresentation of QPE host galaxies among PSB and QBS galaxies is linked to the overrepresentation for TDEs, it is worth investigating if there is any evidence that dust may obscure recent star formation in QPE host galaxies.

The distribution of dust attenuation among the QPE host galaxies and the control galaxies can be seen in the lower panel of Figure~\ref{fig:bagpipes_results}. A Kolmogorov–Smirnov (KS, \citealt{Kolmogorov, Smirnov}) test determines that we cannot distinguish the dust distributions of the QPE hosts from the control galaxies (statistic = 0.32, \textit{p}-value = 0.26). To test whether or not there is a “missing” population of dusty QPE host galaxies at young burst ages, we divide the galaxies into a “young burst” sample and an “old burst” sample, split at a burst age of $10^{-0.5}$ Gyr, the approximate midpoint of the burst ages. We then performed two-sample KS tests on dust attenuation values of the “young burst” and “old burst” QPE host galaxies, and the “young burst” control galaxies and “young burst” QPE host galaxies. None of the resulting \textit{p}-values are significant\footnote{The KS test between the “young burst” and “old burst” QPE host galaxies returns a statistic of 0.6 and a \textit{p}-value of 0.36, while the KS test between the “young burst” control galaxies and “young burst” QPE host galaxies returns a statistic of 0.35 and a \textit{p}-value of 0.51.}, which suggests dust does not disproportionally obscure QPE host galaxies with recent bursts of star formation. 

An additional concern is potential contamination in the QPE host galaxy spectrum from UV-bright TDE disks \citep{vanVelzen_2019, Mummery_2024}, even when the TDE has occurred more than a year before the date of the spectrum, as multiple QPEs have occurred promptly after TDEs and may require TDE disks to exist in the first place. \cite{Newsome_2025} find the late-time UV contribution from a TDE disk to be $\sim$10\% in a spectrum of the host of ASASSN-14li, covering the innermost portion of the nucleus. The relative contribution would decrease when using a spectrum with a larger aperture size, as we do here. \cite{Shepherd_2026} conclude in Section 6.2.4 that the contribution of the UV plateau arising from TDE disks to the host spectra is likely smaller than the calibration uncertainty in \textsc{Bagpipes}'s spectral fit. 

As shown in Appendix~\ref{appendix:a}, the burst mass fraction and burst age can be sensitive to the chosen functional form of the SFH, so the location of the peak in the DTD may depend on the choice of SFH parameterization. However, the QPE-versus-TDE host burst mass fraction comparison is robust to our choice of SFH parameterization because the same method is used in \cite{Shepherd_2026} to determine which host galaxies have high burst mass fractions. This comparison should be unaffected by systematic trends.

\subsection{Implications for QPE Formation Channels}
\label{sec:implications}

Multiple pieces of evidence link QPEs to TDEs, and one of the leading theories of QPE formation channels requires an EMRI to repeatedly intersect with a recently-formed TDE disk. Other QPE formation theories invoke AGN disk instabilities, repeating partial TDEs, or interacting stellar EMRIs. If QPEs were the result of AGN disk instabilities or AGN-EMRI interactions, we would not expect to find an overrepresentation of QPEs in PSB or QBS hosts. PSB and QBS galaxies are a small fraction of AGN host galaxies \citep{Zabludoff_2021}, and changing look AGN are not overrepresented in PSB or QBS galaxies \citep{Dodd_2021, Verrico_2025}. However, if QPEs occur from a combination of TDE and AGN accretion disks, we would expect to find QPEs following a diluted form of the host galaxy trends seen for TDEs. 
 
While we do not have a sufficiently large sample to test whether the PSB overrepresentation differs between TDEs and QPEs, we do observe an important difference between the host galaxy properties of TDEs and QPEs. The DTDs of TDEs and QPEs are similar with respect to burst age, but we find evidence that the distribution of burst mass fractions among the two transient samples is different. The fraction of QPE hosts that have high burst mass fractions is significantly higher than the fraction of TDE hosts that have high burst mass fractions. These results tentatively suggest that the strength of the host galaxy's burst may be more important to QPE formation than the age of the burst.

The formation of EMRIs may be particularly dependent on the SFH of the nucleus of the galaxy. \cite{Naoz_2022} investigate the combined effects of 2-body relaxation and the eccentric Kozai-Lidov mechanism on EMRI rates. Specifically, EMRI rates may be enhanced in systems with a supermassive black hole (SMBH) binary, and \cite{Naoz_2022} suggest that PSB galaxies may have enhanced EMRI rates because they are likely candidates for a SMBH binary arising from a merger. \cite{Metzger_2022} propose that interacting stellar EMRIs cause QPEs, negating the need for a TDE-caused disk but increasing the preference for EMRIs. This preference for EMRIs could arise from PSB galaxies that have experienced a major merger, mimicking TDE host galaxy properties. Perhaps having a strong burst of star formation in a galaxy boosts the TDE rate \textit{and} EMRI rate, and if QPEs require a TDE and an EMRI, then QPE host galaxies may exhibit a strong preference for a strong burst of star formation.

The observational TDE DTD presented in \cite{Shepherd_2026} disagreed with theoretical DTDs from \cite{Stone_2016, Bortolas_2022, Melchor_2024, Wang_2024}, and \cite{Teboul_2025}. To test any disagreement between the observational TDE DTD presented and theoretical DTDs, \cite{Shepherd_2026} performed inverse transform sampling. Due to the larger sample size of that study, they were able to obtain a significant result. We now apply the same test to the observational QPE DTD and theoretical TDE DTDs. We assume one of the theoretical TDE DTDs is correct and then draw theoretical QPE host burst ages from that distribution. We fit a functional form to the $\gamma$=2.25 strong scattering model from \cite{Teboul_2025}, chosen as a representative example of a declining DTD. We draw seven burst ages from this model (representing seven potential high burst mass fraction QPE hosts) and compare this burst age distribution to 100 burst ages from a flat distribution (representing 100 potential control galaxies). Using the Anderson–Darling two-sample test, we find that $\sim$12\% of the time, the sample of burst ages from simulated QPE hosts and the sample of burst ages from simulated control galaxies are drawn from different populations (\textit{p}-value $<$ 0.01), and the median QPE host burst age is younger than the median control galaxy burst age. We find the opposite scenario (median QPE host burst age is older than the median control galaxy burst age when \textit{p}-value $<$ 0.01) to be true 0\% of the time out of 1000 tests. The remaining 88\% of the time, there is no significant difference between the distribution of burst ages for the sample of simulated QPE hosts and the sample of simulated control galaxies, as in our results. Thus, unlike the TDE DTD \citep{Shepherd_2026}, we cannot rule out a DTD that decreases as a function of burst age. A larger sample of QPEs is needed to further test differences between the TDE DTD and the QPE DTD, and to further constrain the rate enhancement mechanisms.

\section{Conclusion}
\label{sec:conclusion}

In this paper, we analyzed the optical spectra of a sample of QPE host galaxies to determine any dependence of the QPE rate on the SFHs of the host galaxies. We calculate the degree to which PSB and QBS galaxies are overrepresented among QPE host galaxies and discuss the constraints that QPE host galaxies' SFHs place on QPE formation channels. Our main conclusions are:

\begin{itemize}
  \item The observational QPE DTD, constructed from the subset of QPE host galaxies that have high burst mass fractions and a sample of control galaxies, increases with burst age until it peaks at $\sim$1 Gyr, then decreases. However, our sample size is not sufficient to distinguish any trends from a DTD that is flat over burst age.
  
  \item QPEs are overrepresented among host galaxies that have recently had a strong burst of star formation. The fraction of the QPE host galaxy sample with high burst mass fractions is significantly higher than the fraction of the TDE host galaxy sample with high burst mass fractions (70\% vs. 35.7\%).
  
  \item PSB and QBS galaxies are both overrepresented in our sample, albeit with large uncertainties. PSB (QBS) galaxies are overrepresented among QPE host galaxies by a factor of 49.0$\pm$46.5 (8.6$\pm$5.4) when compared to galaxies in SDSS DR8. PSB (QBS) galaxies are overrepresented among QPE host galaxies by a factor of 27.4$\pm$26.2 (6.7$\pm$4.3) when compared to the stellar mass-matched and redshift-matched expanded control sample. 

\end{itemize}

TDE and QPE host galaxies are found to share some similarities, though the degree of overlap will have to be studied further with an expanded sample. The observations of stronger burst mass fractions in the QPE hosts relative to TDE hosts and control galaxies implies that, if EMRIs are necessary for QPEs to be observed, the EMRI rate may also depend on the host galaxy SFH. Further study of QPEs can lead to estimates of the EMRI rate and their contribution to the stochastic gravitational wave background \citep{Black_2026}.

\section{Acknowledgments}
\label{sec:acknowledgments}

We thank the anonymous reviewer, whose comments and suggestions improved this work.

The authors thank Thomas Wevers and Riccardo Arcodia for providing QPE host galaxy spectra via private communication.

M.S. and K.D.F. acknowledge support from NSF grant AST 22-06164. M.S. thanks the LSST-DA Data Science Fellowship Program, which is funded by LSST-DA, the Brinson Foundation, the WoodNext Foundation, and the Research Corporation for Science Advancement Foundation; her participation in the program has benefited this work. J.T.H. acknowledges support from NASA through the NASA Hubble Fellowship grant HST-HF2-51577.001-A, awarded by STScI. STScI is operated by the Association of Universities for Research in Astronomy, Incorporated, under NASA contract NAS5-26555. F. acknowledges support from NSF AAG grant 2307375. S.M. acknowledges support from grants NASA ADAP 80NSSC24K0666 and NASA NuSTAR data analysis funding 1729326. M.E.V. acknowledges support from NSF grant NSF-2307375.

Funding for SDSS-III has been provided by the Alfred P. Sloan Foundation, the Participating Institutions, the National Science Foundation, and the U.S. Department of Energy Office of Science. The SDSS-III web site is http://www.sdss3.org/.

SDSS-III is managed by the Astrophysical Research Consortium for the Participating Institutions of the SDSS-III Collaboration including the University of Arizona, the Brazilian Participation Group, Brookhaven National Laboratory, Carnegie Mellon University, University of Florida, the French Participation Group, the German Participation Group, Harvard University, the Instituto de Astrofisica de Canarias, the Michigan State/Notre Dame/JINA Participation Group, Johns Hopkins University, Lawrence Berkeley National Laboratory, Max Planck Institute for Astrophysics, Max Planck Institute for Extraterrestrial Physics, New Mexico State University, New York University, Ohio State University, Pennsylvania State University, University of Portsmouth, Princeton University, the Spanish Participation Group, University of Tokyo, University of Utah, Vanderbilt University, University of Virginia, University of Washington, and Yale University.

\software{Astropy \citep{astropy2013, astropy2018, astropy2022}, Matplotlib \citep{matplotlib}, NumPy \citep{numpy}, \textsc{Bagpipes} \citep{Carnall_2018, Carnall_2019}, pyLick \citep{pyLick}, SciPy \citep{2020SciPy-NMeth}}

\bibliography{Bibliography}

\begin{appendices}

\section{Comparison To Previous Work}
\label{appendix:a}

While this study uses a double power law function to parameterize the burst of star formation, other studies \citep{French_2018} have used an exponential function instead. This choice in \cite{French_2018} was motivated by simulations of merging galaxies, which predict that the merger triggers one \citep{Snyder_2011, Hayward_2014} or two \citep{Mihos_1994, Cox_2008, Renaud_2015} exponentially declining bursts of star formation. However, as we fit a sample of galaxies here and in \cite{Shepherd_2026} that are not necessarily confirmed PSB galaxies, we need a more flexible SFH parameterization. The advantage of using a double power law function to model a burst is that the steepness of the decline is a free parameter in \textsc{Bagpipes}, which allows us to identify non-bursty galaxies with ongoing star formation.

\cite{Wevers_2024} present the \textsc{Bagpipes} results of five of the QPE hosts also presented in this work, and we have acquired the results of two other QPE hosts analyzed by Wevers et al. via private communication. The \textsc{Bagpipes} fit instructions from \cite{Wevers_2024} use a two component star formation history, where the older component of star formation is modeled by a delayed exponential function and the new component of star formation is modeled by an exponential function. \cite{Wevers_2024} restrict their burst age to be no older than 2 Gyr, whereas we allow our burst (modeled by a double power law function) to be no older than 3 Gyr. The spectra used in \cite{Wevers_2024} are from MUSE and have an aperture size of 0.5 arcseconds, while only some of the spectra we use are from MUSE and have an aperture size of 2 arcseconds. We compare the \textsc{Bagpipes} results from \cite{Wevers_2024} to our own results for the seven QPE hosts we have in common between our samples in Figure~\ref{fig:comparisons}. A similar comparison between using a double power law function and using an exponential function to parameterize the burst in TDE host galaxies exists in Figure A1 in \cite{Shepherd_2026}.

To disentangle whether the difference in \textsc{Bagpipes} results is due to the difference in fit instructions or the difference in the spectra aperture size, we compare results where the spectra aperture size remains the same and only the fit instructions change (top row), and where the fit instructions stay the same and the spectra aperture size changes (bottom row). We do this for three different quantities: burst age (left column), burst mass fraction (middle column), and dust (right column). Additionally, though some of the spectra used for QPE hosts' \textsc{Bagpipes} results are not from MUSE (another source had a higher quality spectrum), all of the results in this appendix originate from MUSE spectra. In the top row, we see that using a double power law to model the burst of star formation in \textsc{Bagpipes} produces older bursts, higher burst mass fractions, and more dust. In the bottom row, we see that using a spectrum with a larger aperture size overall produces smaller burst mass fractions and more dust. 

\setcounter{figure}{0}
\renewcommand{\thefigure}{\Alph{section}\arabic{figure}}

\begin{figure*}
\begin{center}
\includegraphics[width=0.32\linewidth]{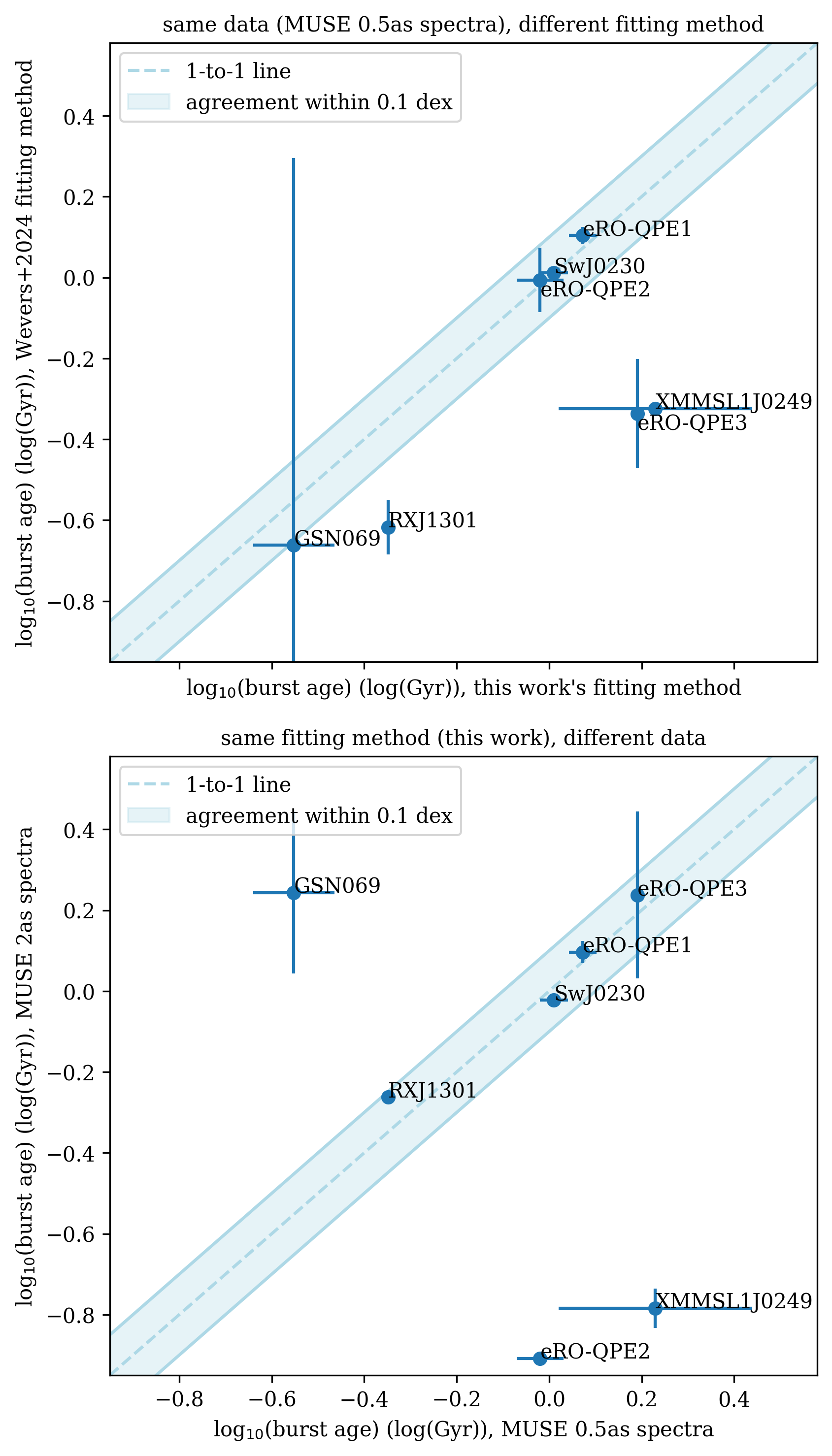}
\includegraphics[width=0.32\linewidth]{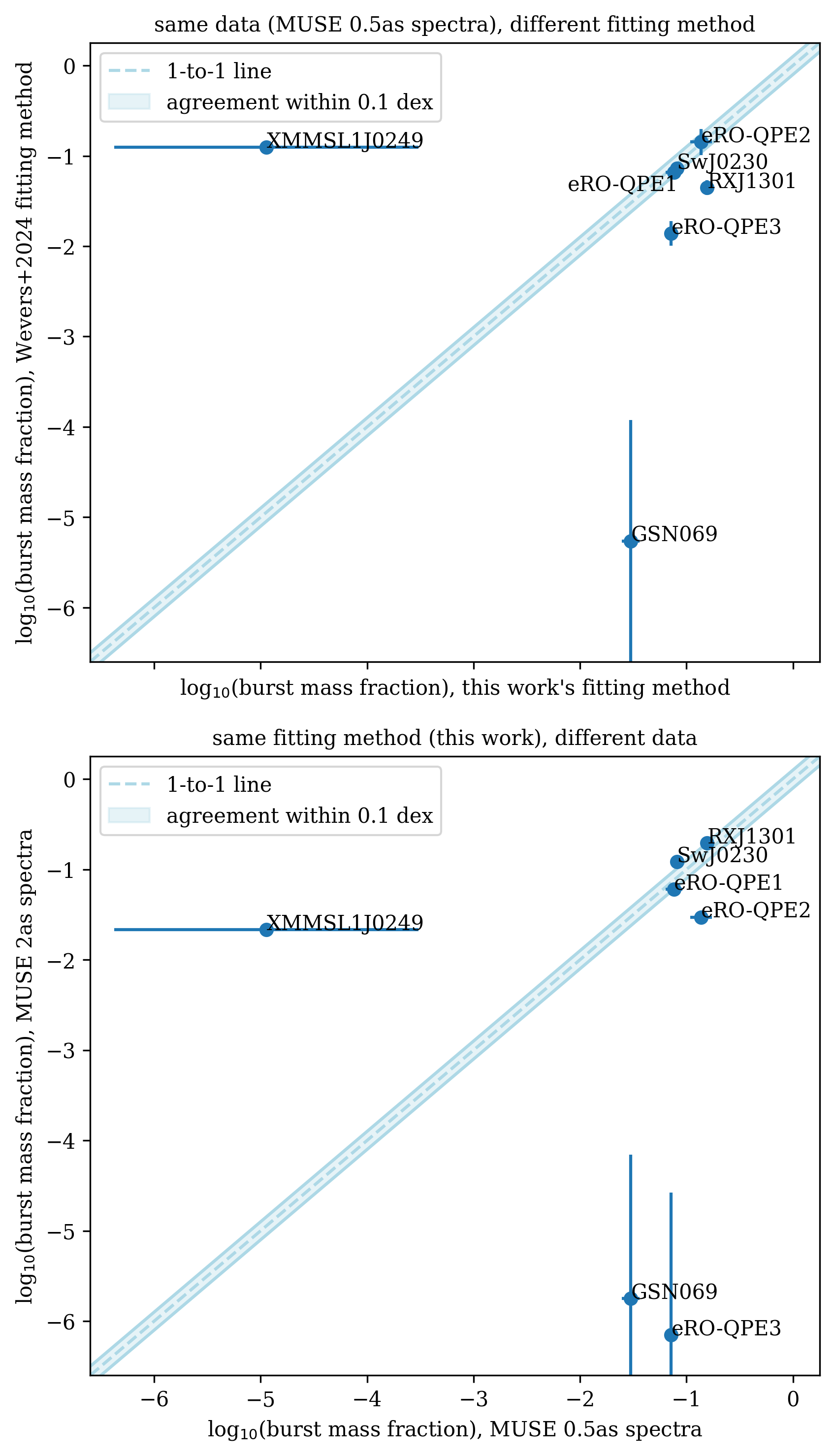}
\includegraphics[width=0.32\linewidth]{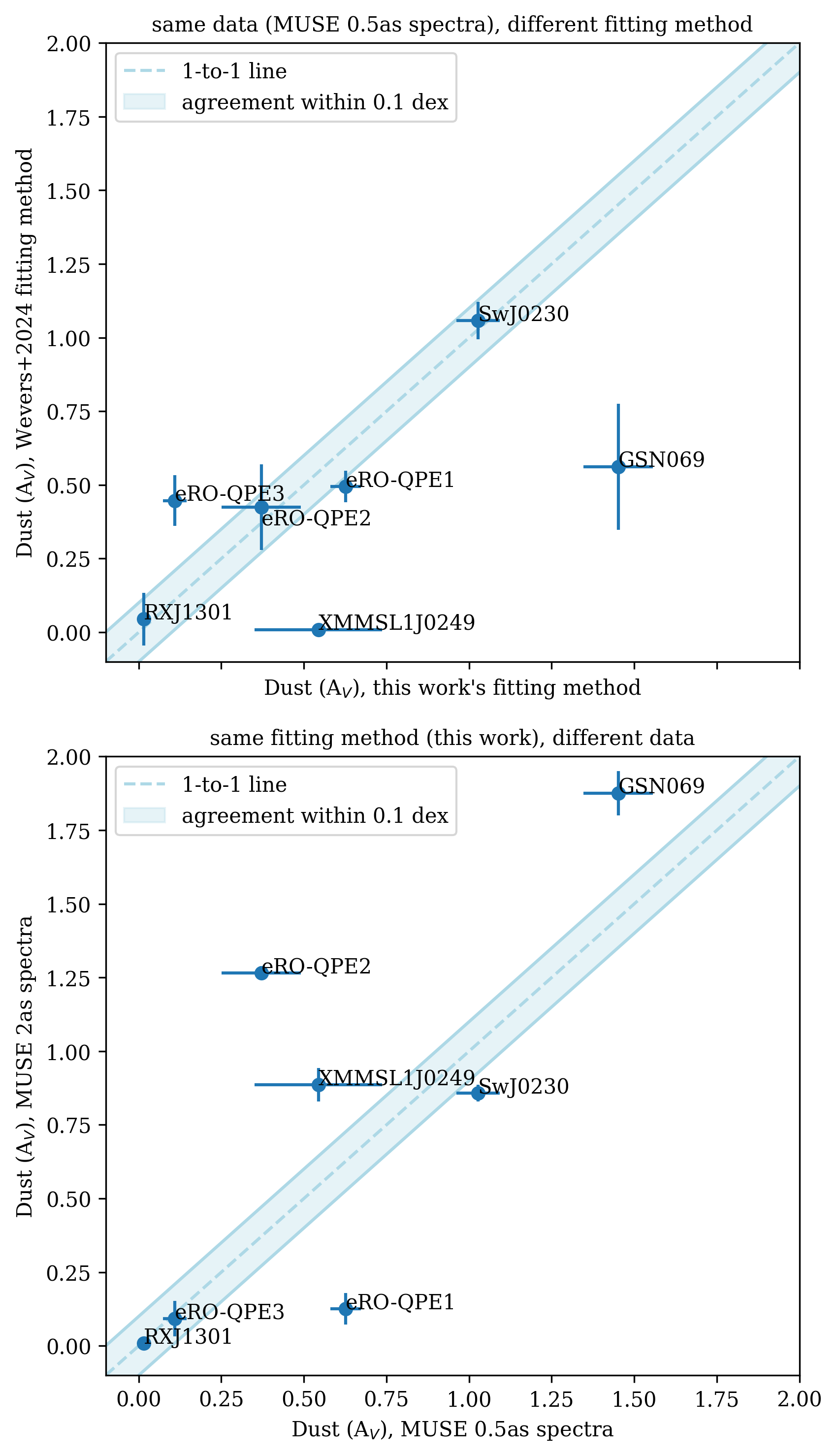}
\end{center}
\caption{Comparison between the \textsc{Bagpipes} results for the burst age \textit{(left column)}, burst mass fraction \textit{(middle column)}, and dust attenuation \textit{(right column)}
of seven QPE hosts in common between our sample and the sample in \cite{Wevers_2024}. \textit{Top row:} Comparison between results using different fit instructions in \textsc{Bagpipes}, where the spectra aperture size is the same. \textit{Bottom row:} Comparison between results using different spectra aperture sizes, where the \textsc{Bagpipes} fit instructions are the same. Deviations away from the one-to-one line in each plot show whether the difference in results is due to the difference in fit instructions or aperture size. For example, the difference in the burst age in the host of eRO-QPE3 is due to the different fitting method, where using the double power law to model the burst produces an older burst age.}
\label{fig:comparisons}
\end{figure*}

\section{Table of Host Galaxy Parameters and \textsc{Bagpipes} Results}
\label{appendix:b}

Information on the QPE hosts, the spectra used for analysis, \textsc{Bagpipes} results, and more can be found in a machine-readable table in the online journal. Table~\ref{tab:megatable} provides a list of columns and descriptions.

\setcounter{table}{0}
\renewcommand{\thetable}{\Alph{section}\arabic{table}}

\begin{table*}[t]
  \centering
  \begin{tabular}{ll}
    \hline
    \hline 
    Column Name & Column Description \\
    \hline
    \hline 
    QPE name & Name of the QPE \\
    \hline
    Host galaxy RA & Right ascension of the host galaxy (degrees) \\
    \hline
    Host galaxy Dec & Declination of the host galaxy (degrees) \\
    \hline
    Host galaxy $z$ & Redshift of the host galaxy \\
    \hline
    Host galaxy $M_{\odot}$ & Stellar mass of the host galaxy ($\log_{10}(M_*/M_{\odot})$) \\
    \hline
    $M_{\odot}$ source & Source of the stellar mass \\
    \hline
    Spectrum source & Source of the spectrum \\
    \hline
    Aperture size & Aperture size or slit width used to take the spectrum (arcseconds) \\
    \hline
    QPE discovery paper & Citation of the QPE discovery \\
    \hline
    Host galaxy H$\alpha$ index & H$\alpha$ index (\AA), positive values indicate lines in emission \\
    \hline
    Host galaxy H$\alpha$ index uncertainty & Uncertainty of the H$\alpha$ index (\AA) \\
    \hline
    Host galaxy H$\alpha$ index source & Source of the H$\alpha$ index \\
    \hline
    Host galaxy H$\delta_{\rm A}$ index & H$\delta_{\rm A}$ index (\AA), positive values indicate lines in absorption \\
    \hline
    Host galaxy H$\delta_{\rm A}$ index uncertainty & Uncertainty of the H$\delta_{\rm A}$ index (\AA) \\
    \hline
    Host galaxy H$\delta_{\rm A}$ index source & Source of the H$\delta_{\rm A}$ index \\
    \hline
    Galaxy classification & Post-starburst (PSB), quiescent Balmer-strong (QBS), star forming (SF), or Quiescent \\
    \hline
    $t_{burst}$ & Time since a burst of star formation in the host galaxy (Gyr) \\
    \hline
    $\sigma_t$ & Standard deviation of $t_{burst}$ (Gyr) \\
    \hline
    $M_{*, burst}$ & Stellar mass formed in the burst ($\log_{10}(M_*/M_{\odot})$) \\
    \hline
    $\sigma_{M_{*, burst}}$ & Standard deviation of $M_{*, burst}$ ($\log_{10}(M_*/M_{\odot})$) \\
    \hline
    $M_{*, old}$ & Stellar mass formed in the old stellar component ($\log_{10}(M_*/M_{\odot})$) \\
    \hline
    $\sigma_{M_{*, old}}$ & Standard deviation of $M_{*, old}$ ($\log_{10}(M_*/M_{\odot})$) \\
    \hline
    $M_{*, burst}/M_{*, tot}$ & Burst mass fraction \\
    \hline
    $\alpha$ & Falling slope index of the double power law (burst) component \\
    \hline
    $\sigma_{\alpha}$ & Standard deviation of $\alpha$ \\
    \hline
    $A_V$ & Dust \\
    \hline
    $\sigma_{A_V}$ & Standard deviation of $A_V$ \\
    \hline
  \end{tabular}
  \caption{Column names and column descriptions for the ancillary information and \textsc{Bagpipes} results pertaining to the QPE host galaxies. The values in the $t_{burst}$, $M_{*, burst}$, $M_{*, old}$, $\alpha$, and $A_V$ columns are the median values of the posteriors returned by \textsc{Bagpipes}. This table is available in its entirety in a machine-readable form in the online journal.}
  \label{tab:megatable}
\end{table*}

\end{appendices}

\end{document}